# From the Top Down: does Corruption affect Performance?

MAURIZIO LA ROCCA[*] TIZIANA LA ROCCA[•]
FRANCESCO FASANO[♠] F. JAVIER SANCHEZ-VIDAL[**]

**Abstract**

Corruption, fraud, illicit and unethical activities have become a worldwide major impediment to economic, political, and social development. While existing empirical analyses are mainly based on measures of corruption at country level, this is the first empirical research work, based on a large international dataset, to measure the impact of illicit, and unethical activities by managers and test their consequences for firm performance. Using cross-sectional data, the aim of this study is to analyze the impact of corruption on firm performance. Our definition of corruption considers whether and to what extent managers are involved in mismanagement, misconduct or corruption. The implications for corporate governance when it comes to the managers' appointment procedure are as vital as strategic.

### Framing of the research.

A government official, bureaucrat, or politician is said to have engaged in public corruption when they are involved in unethical behavior that includes offering or accepting financial or non-financial benefits from other government officials or private individuals. In contrast, private corruption is described as any dishonest activity carried out by a worker, manager, or organization that entails the provision of benefits to other private or public individuals or organizations. It is undeniable that corruption is becoming more and more relevant; in 2018, UN Secretary-General António Guterres proclaimed it to be a global issue that costs the equivalent of 5% of global GDP and denies communities access to public services that promote growth, such as hospitals and schools. Additionally, corruption misallocates entrepreneurial abilities and deters foreign direct investment. In 2022, the Secretary-General reaffirmed his commitment to "address corruption and illicit financial flows" with governments and organizations. "If all countries were to reduce corruption in a comparable manner, they could gain $1 trillion in lost tax revenues, or 1.25% of global GDP," according to Mauro et al. (2019, p. 27). The 2020 Davos manifesto encouraged businesses to adopt a policy of "zero tolerance" for corruption as a result. Dyck et al. (2021) estimate that business fraud reduces equity value by 1.7% annually.

Being such a multifaceted topic, works on corruption can be found in fields as diverse as law (Mijares, 2015), finance (Pantzalis et al. 2008), economics (He et al. 2015), accounting (Everett et al., 2007), and international business (Cuervo-Cazurra, 2016), among others. Corruption has a significant impact on sociology, law, political science, economics, and management. Although "corruption" is usually associated with the public sphere, the idea also applies to private interactions between businesses. Transparency International publishes reports on various topics where corruption is important, including the private sector, which can struggle to be a driver of innovation and development because of the destabilizing impact that corruption has on fair competition.

---

[*]   Associate Professor *(Management and Finance)* – University of Calabria (Italy)
      e-mail: _maurizio.larocca@unical.it_
[•]   Associate Professor *(Management)* – University of Messina (Italy)
      e-mail: __ elviratiziana.larocca@unime.it __
[♠]   Assistant Professor *(Management and Finance)* – University of Calabria (Italy)
      e-mail: __ francesco.fasano@unical.it __
[**]  Associate Professor *(Finance)* – Universidad Politécnica de Cartagena (Spain)
      e-mail: __ javier.sanchez@upct.es __

Additionally, the so-called relational capital, or social connections that managers and organizations make with important market participants like other businessmen and government officials to support value creation processes, may be used in an unethical and unfair way. Managers and business owners, for instance, might prefer to invest their time and resources in strategic development projects to compete in the market than in the search for favoritism in the allocation of permits, licenses, and government funds.

Because we want to analyze both ethical and unethical corporate behaviors, we will refer to corporate corruption and mishandling in this work rather than corporate crime. The analysis of legal but unethical corporate practices has been neglected, despite the fact that the majority of literature uses the terms fraud and corporate corruption interchangeably and analyzes cases of corruption between individuals inside organizations associated with criminal practices (Beasley et al., 2000). Ahmed (2020) uses a broad definition of corruption that includes "a wide range of immoral activities, including, but not limited to, bribery, nepotism, extortion, cronyism, embezzlement, lobbying, rent-seeking, state capture, patronage, revolving door phenomenon, clientelism, partisanship, and tax evasion".

The following ideas were used as keywords in a recent literature survey on corruption by Bahoo et al. (2020): Abuse, Crime, Criminal, Decay, Extortion, Falsification, Fraud, Graft, Manipulation, Misconduct, Misrepresentation, and Wrongdoing are all examples of corruption. This, in our opinion, provides a broad enough definition of the phenomenon that this work refers to as corruption. The percentage of managers who reported incidents of misconduct or poor management will be used as a proxy for these issues. We do not want to restrict the study of the relationship between corruption and firm performance to bribery, as this would exclude phenomena like nepotism, cronyism, etc., as De Rosa et al. (2010) point out. Criminal histories of CEOs and lower-level workers are used by Regenburg and Bigler Seitz (2021), who discover a positive correlation between these histories and a company's likelihood of bankruptcy. Bianchi et al. (2022) used managers' criminal records showing involvement with criminal organizations and found a negative relation with performance.

Due to its illegality and secrecy, business corruption is difficult to measure. According to Dyck et al. (2010), corporate fraud is more frequently reported by employees (17% of the time), by non-financial authorities or institutions (13%), or by the media (13%), rather than by members of the BvD, who would be expected to report it. This highlights the inability of internal control mechanisms to identify and prevent widespread corporate scandals. One might draw the conclusion that supervisors frequently participate in these unethical practices and are complicit in them.

While there have been a few contributions at the firm level that examine the impacts on corporate performance, most of the literature on the topic of corruption focuses on macroeconomic issues. Two lines of inquiry can be distinguished in the analysis of how corruption affects businesses and the economy: the "sand the wheels" argument and the "grease the wheels" viewpoint. According to the former, graft hinders economic development, discourages investment, and creates trade frictions (Shleifer and Vishny, 1993; Mauro, 1995; Kaufmann and Wei, 1999; Mo, 2001; Méon and Sekkat, 2005; Svensson, 2005; Fisman & Svensson, 2007; De Rosa et al., 2015). The latter presupposes that corruption causes effective bureaucracy to decrease; thus, corruption could serve as a lubricant to counteract the flaws brought on by a rigid and bureaucratic system that restricts business activity (Leff, 1964; Huntington, 1968; Acemoglu and Verdier, 1998; Kaufmann and Wei, 1999: Vial and Hanoteau, 2010; Krammer, 2019).

Prior studies at the micro (firm) level have frequently concentrated on business survey data or perception survey data at the national and aggregated level, such as the World Bank indexes: Corruption Control Index (CCI) or Corruption Perception Index (CPI), which aim to measure how the level of corruption at a country level is supposed to influence firm performance; and the use of surveys of people inside and outside of firms. The disadvantages of both approaches are clear.

A firm-level corruption measure that considers the level of corruption of each manager in each unique business shows interesting and innovative results for an analysis of the effects of corporate corruption on performance. Since firm heterogeneity probably affects firm performance, we avoid the drawback of using aggregated data that cannot account for it (Kasahara and Rodrigue, 2008). Due to the potential for interviewee distortion caused by dishonest answers (Kaufmann and Wei, 1999), this type of measure has an additional advantage over commonly used measures of perception of corruption based on surveys. As far as we are aware, there are no studies that directly assess the degree of managers' involvement in corporate corruption.

Regarding the body of research on corporate corruption, some studies examined its effects on corporate growth (e.g., Tanzi and Davoodi, 2000; Fisman and Svensson, 2007; Kimuyu, 2007; Wang and You, 2012; Ayaydn and Hayaloglu, 2014), while others examined its effects on company performance, despite basing the measure of corruption on surveys and perceptions, either at the regional or the national level (Gaviria, 2002; McArthur and Teal, 2002; Donadelli et al., 2014; Williams et al., 2016; Martins et al., 2020).

According to some research, performance and corruption have a positive relationship. Ayaydn and Hayaloglu (2014) demonstrate the presence of a statistically significant and positive relationship between the degree of corruption and business development using a relatively small sample of 41 manufacturing companies from Turkey. The authors contend that by promoting trade, corruption, as assessed by a perception index, quickens business economic growth. Williams et al. (2016) find that corruption improves the company's performance, as measured by growth in turnover and employment above and beyond that from productivity, using a sample of firms from 132 developing countries for the years 2006–2014 and data from the World Bank Enterprise Survey (WBES). According to Ashyrov and Akuffo (2020), political inefficiency and bureaucracy increase company productivity.

But the majority of studies indicate a detrimental impact on corporate success. Gaviria (2002) evaluates the effects of corruption and crime on sales, investments, and company development using data from a survey of Latin American businesses, showing a strong negative effect on company competitiveness. Focusing on the African setting, McArthur and Teal (2002) emphasize how corporate corruption has a detrimental impact on a company's output. Athanasouli et al. (2012) demonstrate how administrative corruption has a "commercial barrier" that impairs the performance of the firm by concentrating on a sample of Greek businesses and Business Environment and Enterprise Performance Survey (BEEPS) data. Office abuse exhibits a negative effect on the company's productivity, according to De Rosa et al. (2010) who used data from a BEEPS survey of 11,000 businesses in 28 developing and developed countries. Faruq et al. (2013) examined the effects of bribes, nepotism, covert funding of political parties, and suspected ties between politics and business on the productivity of 900 companies over a twelve-year span using data from Ghana, Kenya, and Tanzania, three African nations. Their findings indicate that corruption and inefficient bureaucracy have a significant detrimental effect on company productivity. Van Vu et al. (2018) demonstrate that the degree of corruption has a statistically significant and negative effect on ROA using a national survey with institutional quality data (at the provincial level) and a sample of manufacturing SMEs. Nguyen et al. 2022 also discovered this inverse relationship. According to Bianchi et al. (2022), managers who may have ties to illegal groups are less profitable. According to research by Yang et al. (2021), foreign companies' financial success is negatively correlated with the degree of national corruption.

### Theoretical basis.

It is reasonable to assume that the company managers' involvement in illegal, dishonest, and unethical activities harms the operation and management of the company and lowers its efficiency in light of the prevailing literature, which typically detects a negative effect of corporate corruption on performance. Due to payments for bribes and various bonuses, these businesses will incur higher operating costs in addition to reputational costs. Previous research has noted that because corruption "permeates the critical aspects of a state's organization," mismanagement and corruption, when present at high levels, can have a domino impact (Ashforth and Anand 2003). Similar proxy measures for corruption have been used in other works; for instance, Butler et al. 2009 or Zhang et al. 2018 use the number of state official convictions per capita for abuse of public office to gauge the level of national corruption.

It can be assumed that the company managers' involvement in illicit, dishonest, and unethical activities harms the operation and management of the company, reducing its efficiency, in light of the prevailing literature, which typically detects a negative effect of corporate corruption on performance. Payments for bribes and different bonuses will result in higher operating costs for these businesses, along with reputational costs. As corruption "permeates the essential aspects of a state's organization," previous literature has noted that corruption and mismanagement, when present at high levels, can have a domino effect (Ashforth and Anand 2003). Similar proxies for misconduct have been used in other works; the number of state official convictions for abuse of public office per capita is used by Butler et al. 2009 or Zhang et al. 2018 to gauge the degree of national corruption.

The secretive nature of corrupt or dishonest behavior interrupts, or at the very least interferes with, the valuable flow of information and transparency that should characterize any business. It stands to reason that companies with such unethical modus operandi will tend to be less transparent in order to hide such behavior. In this regard, Athanasouli and Goujard (2015) contend that corruption worsens business management by lowering the caliber of managerial decisions, decreasing confidence in company expertise, fostering inefficiencies, lowering R&D and investments, and, ultimately, harming business productivity.

According to agency theory, a dishonest environment could make opportunism worse by encouraging people to seek out opportunities to advance their own interests. Based on the research done by Zahra et al. (2005), it can be assumed that opportunistic people can use company resources for their own benefit and conceal important information from the controllers. This opportunistic behavior exacerbates issues with information inequality and a lack of trust in corporate abilities, leading to higher agency costs. Some managers might exploit the information at their disposal to acquire higher positions of power and personal advantages and conceal crucial information to conceal their dishonest and illegal behavior (Braun and Guston, 2003). (Williamson, 1985). Corrupt practices may lead to a decline in a company's image, culture, the ability to allocate resources effectively, and the drive for innovation (Lou 2002; Hung 2008). As a result, debt holders will be less willing to lend money or will do so at a higher cost, or there may be some customer reluctance that will result in declining sales. We also believe that it will raise the risk perception of the company's key stakeholders. Additionally, vendors will harbor mistrust for the business, resulting in a less agile operational cycle overall. As Sampath et al. (2018) discovered, the reputational damage caused by bribery incidents may be the cause of this.

The story of Enron and its CFO, Andrew Fastow, serves as an example. He was in charge of all the unique, intricate, and financially savvy off-balance sheet activities (limited liability partnership controlled by Enron) used to hide Enron's enormous losses in these affiliated and controlled companies' quarterly financial statements. He participated in Enron's corrupt and fraudulent activities, was aware of them, and retained illegal personal holdings in such fictitious, seemingly independent companies. In this manner, Fastow was accountable for Enron's attempt to defraud its stakeholders while also defrauding Enron of tens of millions of dollars. This is an illustration of how managerial corruption amplifies opportunistic sentiment with a greater detrimental impact on firm performance. The costs of the business increased as a result of the stakeholders' growing agency conflicts and the worsening information asymmetry.

First hypothesis (H1): Performance of the company is adversely impacted by management corruption levels.

**Purpose of the paper.** The purpose of this research paper is to examine through empirical evidence how the degree of corporate corruption, indicated by the level of participation of managers in illicit or unethical behavior, impacts a company's performance. The study aims to determine whether such behavior is advantageous for a company's competitive position against its rivals or whether it is a hindrance to the company's best interests. Our interpretation is consistent with the findings of Davidson et al. (2015), who discovered that executives, specifically CEOs and CFOs, who have a criminal history are more likely to engage in fraudulent activities.

**Methodology.** The methodology is divided into three parts: Sample, Variables, and Methodology. In the Sample section, the Orbis database by Bureau van Dijk (BVD) and LexisNexis are used to analyze the relationship between corruption at the enterprise level and performance. The LexisNexis WorldCompliance database is used to measure the percentage of corrupt executives in the companies. In the Variables section, the return on assets (ROA) (Sánchez-Vidal et al. 2023) is used as the dependent variable, and size, average executive age, board size, board gender diversity, ownership structure, firm age, financial leverage, sales growth, tangibility of corporate assets, R&D intensity, and dummy industry and country variables are used as control variables. In the Methodology section, the model used to test the hypotheses is presented as firm performance = f (corporate corruption, control variables). However, the problem of endogeneity is addressed by applying the two-stage least squares (2SLS) estimation technique.

**Results.** No results yet

**Research limitations.** In future studies examining corruption and financial performance, it would be advisable to take into account time-specific effects by using panel data in conjunction with cross-sectional effects. To further understand the impact of corruption and financial performance, it is also suggested to investigate the moderating effect of corporate social responsibility. It would be valuable to investigate the specific ways in which mismanagement contributes to poor performance, such as considering the costs associated with reputation, investigation, and monitoring.

**Managerial implications.** If there is a negative relation between corruption and performance, managers should always practice ethical and virtuous behaviors and avoid engaging in corrupt actions, even if they are intended to promote the company's interests. The potential benefits of such actions are often outweighed by the related costs, such as increased opportunism and management inefficiencies that result in lower quality decision-making. Policymakers should recognize that to maximize the overall well-being of companies, they must combat corruption since it harms their overall value.

**Originality of the paper.** This study is original because it is the only one that measures the level of corruption at the individual firm level while considering its management and associates it with the overall company performance.